\documentclass[prx,reprint,superscriptaddress,amsmath,amssymb]{revtex4-2}
\usepackage{graphicx}

\begin{document}

\title{Observation of the Electronic Pomeranchuk Effect in Generalized Wigner Crystals of Twisted MoS$_2$}

\author{Xinjie Fang}
\affiliation{Department of Physics, Fudan University, Shanghai 200433, China}
\affiliation{Department of Physics, School of Science, Westlake University, Hangzhou 310024, Zhejiang Province, China}
\affiliation{Institute of Natural Sciences, Westlake Institute for Advanced Study, Hangzhou 310024, Zhejiang Province, China}

\author{Linfeng Wu}
\affiliation{Department of Physics, Fudan University, Shanghai 200433, China}
\affiliation{Department of Physics, School of Science, Westlake University, Hangzhou 310024, Zhejiang Province, China}
\affiliation{Institute of Natural Sciences, Westlake Institute for Advanced Study, Hangzhou 310024, Zhejiang Province, China}

\author{Naitian Liu}
\affiliation{Department of Physics, Fudan University, Shanghai 200433, China}
\affiliation{Department of Physics, School of Science, Westlake University, Hangzhou 310024, Zhejiang Province, China}
\affiliation{Institute of Natural Sciences, Westlake Institute for Advanced Study, Hangzhou 310024, Zhejiang Province, China}

\author{Le Zhang}
\affiliation{Department of Physics, School of Science, Westlake University, Hangzhou 310024, Zhejiang Province, China}
\affiliation{Institute of Natural Sciences, Westlake Institute for Advanced Study, Hangzhou 310024, Zhejiang Province, China}

\author{Wenqiang Zhou}
\affiliation{Department of Physics, School of Science, Westlake University, Hangzhou 310024, Zhejiang Province, China}
\affiliation{Institute of Natural Sciences, Westlake Institute for Advanced Study, Hangzhou 310024, Zhejiang Province, China}

\author{Jing Ding}
\affiliation{Department of Physics, Fudan University, Shanghai 200433, China}
\affiliation{Department of Physics, School of Science, Westlake University, Hangzhou 310024, Zhejiang Province, China}
\affiliation{Institute of Natural Sciences, Westlake Institute for Advanced Study, Hangzhou 310024, Zhejiang Province, China}

\author{Zhangyuan Chen}
\affiliation{Department of Physics, Fudan University, Shanghai 200433, China}
\affiliation{Department of Physics, School of Science, Westlake University, Hangzhou 310024, Zhejiang Province, China}
\affiliation{Institute of Natural Sciences, Westlake Institute for Advanced Study, Hangzhou 310024, Zhejiang Province, China}

\author{Hanxiao Xiang}
\affiliation{Department of Physics, Fudan University, Shanghai 200433, China}
\affiliation{Department of Physics, School of Science, Westlake University, Hangzhou 310024, Zhejiang Province, China}
\affiliation{Institute of Natural Sciences, Westlake Institute for Advanced Study, Hangzhou 310024, Zhejiang Province, China}

\author{Yue Xiao}
\affiliation{Department of Physics, School of Science, Westlake University, Hangzhou 310024, Zhejiang Province, China}
\affiliation{Institute of Natural Sciences, Westlake Institute for Advanced Study, Hangzhou 310024, Zhejiang Province, China}

\author{Kenji Watanabe}
\affiliation{Research Center for Electronic and Optical Materials, National Institute for Materials Science, 1-1 Namiki, Tsukuba 305-0044, Japan}

\author{Takashi Taniguchi}
\affiliation{Research Center for Materials Nanoarchitectonics, National Institute for Materials Science, 1-1 Namiki, Tsukuba 305-0044, Japan}

\author{Shuigang Xu}
\email{xushuigang@westlake.edu.cn}
\affiliation{Department of Physics, School of Science, Westlake University, Hangzhou 310024, Zhejiang Province, China}
\affiliation{Institute of Natural Sciences, Westlake Institute for Advanced Study, Hangzhou 310024, Zhejiang Province, China}

\date{\today}

\begin{abstract}
Moir\'e superlattices in transition metal dichalcogenides provide a
highly tunable platform for exploring strongly correlated electronic
phases, such as generalized Wigner crystals. While these crystalline
states typically melt with increasing thermal fluctuations, an
electronic analogue of the Pomeranchuk effect can stabilize the
localized solid phase at elevated temperature through isospin
entropy. Here, we report the observation of an electronic Pomeranchuk
effect at the fractional filling factors of $\nu = 1/3$ and $\nu = 1/4$
in AB-stacked twisted bilayer MoS$_2$ with twist angles of 4.1$^\circ$
and 3.9$^\circ$, respectively. At ultra-low temperature, the system
exhibits a highly conducting, itinerant behavior at these fractional
fillings, characteristic of a compressible Fermi-liquid ground state.
Upon heating, the system exhibits a counterintuitive increase
in the longitudinal resistivity, signaling an isospin-entropy-driven
transition into a localized generalized Wigner crystal state. Our
findings highlight the unique capacity of flat bands in twisted bilayer
MoS$_2$ for stabilizing highly degenerate magnetic configurations,
offering new insights into the thermodynamic phase diagrams of
low-dimensional correlated systems.
\end{abstract}

\maketitle

\section{Introduction}

The realization of flat electronic bands in two-dimensional van der
Waals moir\'e superlattices has opened a vibrant frontier for studying
strong electron-electron interactions and symmetry-broken states. In
transition metal dichalcogenides (TMDC), the quenched kinetic energy
within these narrow bands allows long-range Coulomb repulsion to
dominate the system's behavior even at fractional fillings of the moir\'e
unit cell [1--7]. Under these conditions, electrons minimize their
electrostatic energy by localizing onto specific sublattices, forming
correlation-driven insulating phases known as generalized Wigner
crystals (GWC) [7--10]. Generally, such crystalline phases are expected
to be most robust at base temperature, undergoing conventional thermal
melting into a disordered Fermi liquid as temperature increases due to
the proliferation of spatial fluctuations. However, thermodynamics
allows for a counterintuitive alternative where heating drives a
transition from a liquid to a solid---a phenomenon classically
exemplified by the Pomeranchuk effect in helium-3 [11,12]. In the
electronic analogue of this effect, the localized solid phase possesses
a substantially larger entropy than the mobile Fermi liquid at low
temperature. In a GWC at low filling fractions such as $\nu = 1/3$, the
spatial separation between localized electrons severely quenches the
magnetic exchange interaction. Consequently, the isospin degrees
of freedom of the localized moments remain highly degenerate and
disordered down to very low temperature, acting as an entropy reservoir
with an entropy approaching $s \approx k_{B}\text{ln}2$ (or
$k_{B}\text{ln}4$ when both spin and valley degeneracies are present)
per localized electron. According to the Clausius-Clapeyron relation,
raising the temperature populates these highly entropic magnetic
configurations, lowering the free energy of the localized phase and
forcing the itinerant electronic fluid to freeze into an insulating
crystal upon heating.

The electronic Pomeranchuk effect has been established in magic-angle
twisted bilayer graphene and certain twisted TMDC at integer moir\'e
fillings [13--16], where the short-range (on-site) Coulomb repulsion
interaction ($U$) serves as the dominant energy scale over the bandwidth
($W$) and governs the Mott insulator (or Wigner crystal) [1,2,7,16].
However, its extension to fractional fillings remains largely unexplored.
At fractional fillings, long-range (inter-site) Coulomb interactions
($V$) are expected to play a more prominent role, potentially
stabilizing charge-density waves, GWC, or fractional topological phases
[3,7,10]. Whether the Pomeranchuk effect can emerge in this regime
remains unknown.

% Figure 1 float is defined here (before its first in-text reference) so
% that it is placed at the top of page 2 rather than page 3.
\begin{figure*}[!t]
\includegraphics[width=\textwidth]{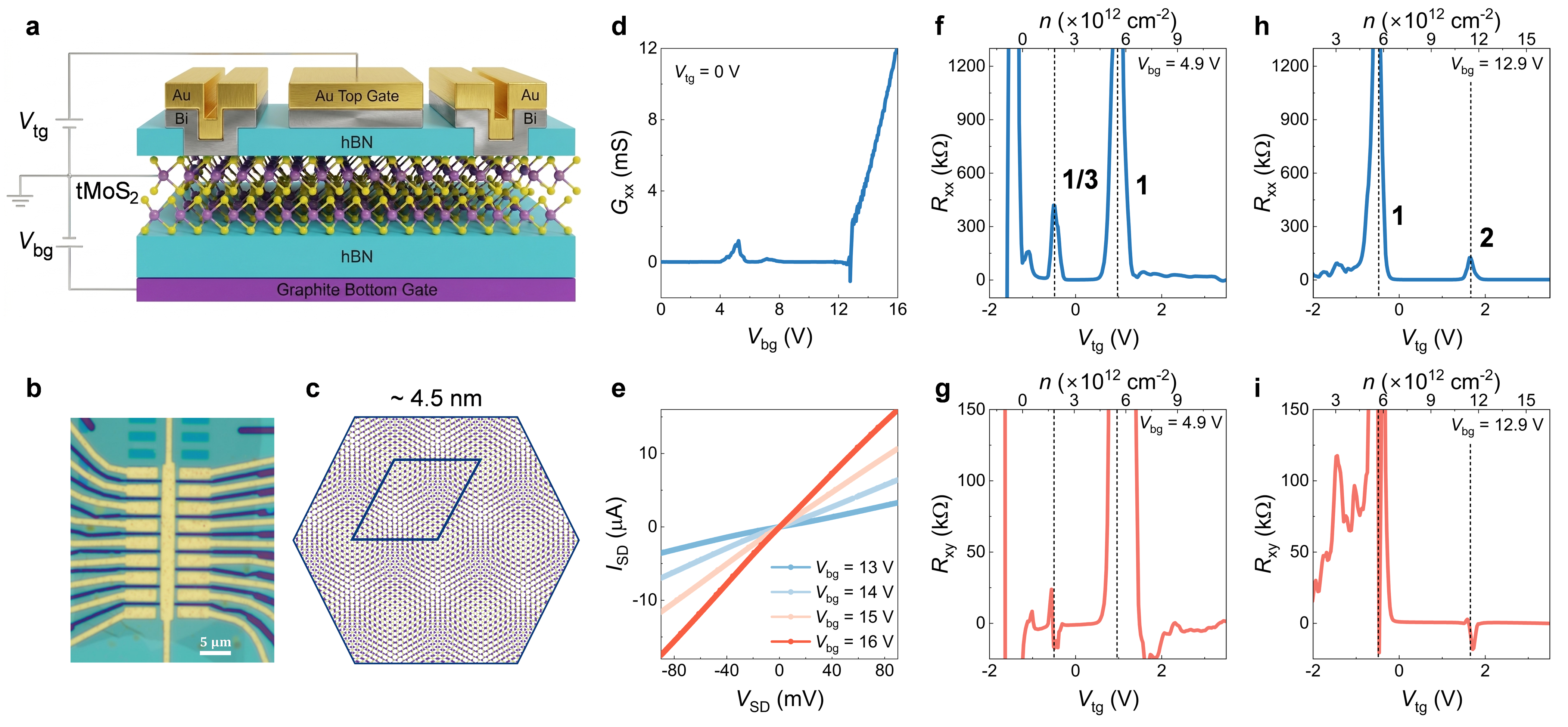}
\caption{Device structure and electrical performance of the tMoS$_2$ FET. \textbf{a}-\textbf{c}, Schematic illustration (a), optical image (b) of the device D1, and depiction of the twisted bilayer MoS$_2$ moir\'e superlattice (c). \textbf{d}, Four-terminal conductance as a function of $V_{bg}$ at fixed $V_{tg} = 0$ V. \textbf{e}, Two-terminal $I_{SD}$-$V_{SD}$ characteristics of a representative tMoS$_2$ device measured at 1.5 K. \textbf{f}, \textbf{h}, Longitudinal resistance $R_{xx}$ plotted as a function of $V_{tg}$ at fixed $V_{bg} = 4.9$ V (f) and $V_{bg} = 12.9$ V (h). \textbf{g}, \textbf{i}, Corresponding transverse resistance $R_{xy}$ plotted as a function of $V_{tg}$, at $V_{bg} = 4.9$ V (g) and $V_{bg} = 12.9$ V (i) under an applied magnetic field $B = 6$ T. The resistance peaks of correlated insulator at $\nu = 1/3$, the Mott insulator at $\nu = 1$, and the band insulator at $\nu = 2$ are marked out.}
\label{fig:1}
\end{figure*}

TMDC moir\'e systems, endowed with multiple internal degrees of freedom
such as spin and valley, provide an ideal platform for entropy-driven
phase transitions, including Pomeranchuk-like crystallization phenomena.
In particular, owing to its robustness under ambient conditions, twisted
bilayer MoS$_2$ (tMoS$_2$) offers a uniquely compelling architecture for
examining this phenomenon at fractional fillings, yet its exploration
remains severely bottlenecked by the technical challenge of fabricating
low-temperature Ohmic contacts [17--26]. In this work, we utilize
temperature-dependent and magneto-transport measurements to
systematically investigate the thermodynamic stability of the
$\nu = 1/3$ and $\nu = 1/4$ GWC states in high-quality AB-stacking
tMoS$_2$ devices. At base temperature, electronic transport
characterizations indicate a highly compressible metallic or
Fermi-liquid ground state at these fractional fillings, which is
corroborated by a quadratic temperature dependence. As the temperature
is increased to around ten Kelvin, we observe an anomalous rise in
the resistivity, demonstrating the robust thermal opening of an
insulating charge gap.

\section{Results and Discussion}

\subsection{High-quality Ohmic contacts and moir\'e band characterization}

Figures 1a and 1b show the schematic and optical image of our dual-gated
device, respectively. The low-work-function semimetal bismuth (Bi)
electrodes were used as the metallic layer to achieve low-temperature
ohmic contacts in tMoS$_2$ [25,26]. As shown in our transport
characterization, the device achieves a remarkably high four-terminal
conductance of $12\text{~mS}$ (Fig.~1d) and a field-effect mobility of
$\mu_{\text{FET}} \approx 4.3 \times 10^{4}\text{~cm}^{2}\ \text{V}^{-1}\text{s}^{-1}$
at $1.5\text{~K}$. The linear two-terminal $I\text{-}V$ curves across a
wide gate voltage range confirm the elimination of Schottky barriers
(Fig.~1e), enabling high-fidelity transport measurements deep into the
cryogenic regime.

The AB-stacked tMoS$_2$ was fabricated by stacking two sheets of
monolayer MoS$_2$ with a twist angle of
$\theta \approx 180^{\circ} + 4^{\circ}$ using the standard cut \& stack
method [27]. The precise twist angle can be directly determined by
low-temperature transport. Specifically, the electron density needed to
fill one electron per moir\'e unit cell ($\nu = 1$) is
$n_{\text{M}} = 2\theta^{2}/\sqrt{3}a^{2}$, where $a = 0.3161$ nm is the
lattice constant of MoS$_2$. From this relation, $\theta$ can be
calculated. In our device D1, we observe
$n_{\text{M}} = 5.8 \times 10^{12}$ cm$^{-2}$ (Figs.~1f-1i),
corresponding to $\theta = {4.1}^{\circ}$ and a moir\'e wavelength of
4.5 nm.

Multiple resistance peaks are observed in longitudinal resistance
$R_{xx}$, along with sign reversals in Hall resistance $R_{xy}$ as shown
in Figs.~1f-1i. The resistive states at $\nu = 2$ and correlated states
at $\nu = 1$ can be attributed to band insulators and Mott insulators,
respectively [1,7]. Additional correlated states appear at fractional
filling as shown in Fig.~1f, which have been previously identified as
the formation of GWC or charge-density waves in other twisted TMDC
systems [3,7,8,10].

\subsection{Multivalley moir\'e band population}

To establish the electronic structure relevant to the fractional-filling
states, we mapped the four-terminal $R_{xx}$ as a function of $\nu$ and
displacement field $D$. Figures 2a and 2b show $R_{xx}(\nu,D)$ at 1.5 K
under perpendicular field $B_{\bot} = 0$ and 11.8 T, respectively. At
$B_{\bot} = 0$ T, the Mott insulating feature at $\nu = 1$ is prominent,
whereas the features near $\nu = 1/3$ and $\nu = 2$ are comparatively
weak. Both become more readily resolved at large $B_{\bot}$. The
persistence of the $\nu = 1$ feature throughout the accessible
displacement-field range indicates that the interaction scale remains
substantial in this device [1].

\begin{figure*}[t]
\includegraphics[width=\textwidth]{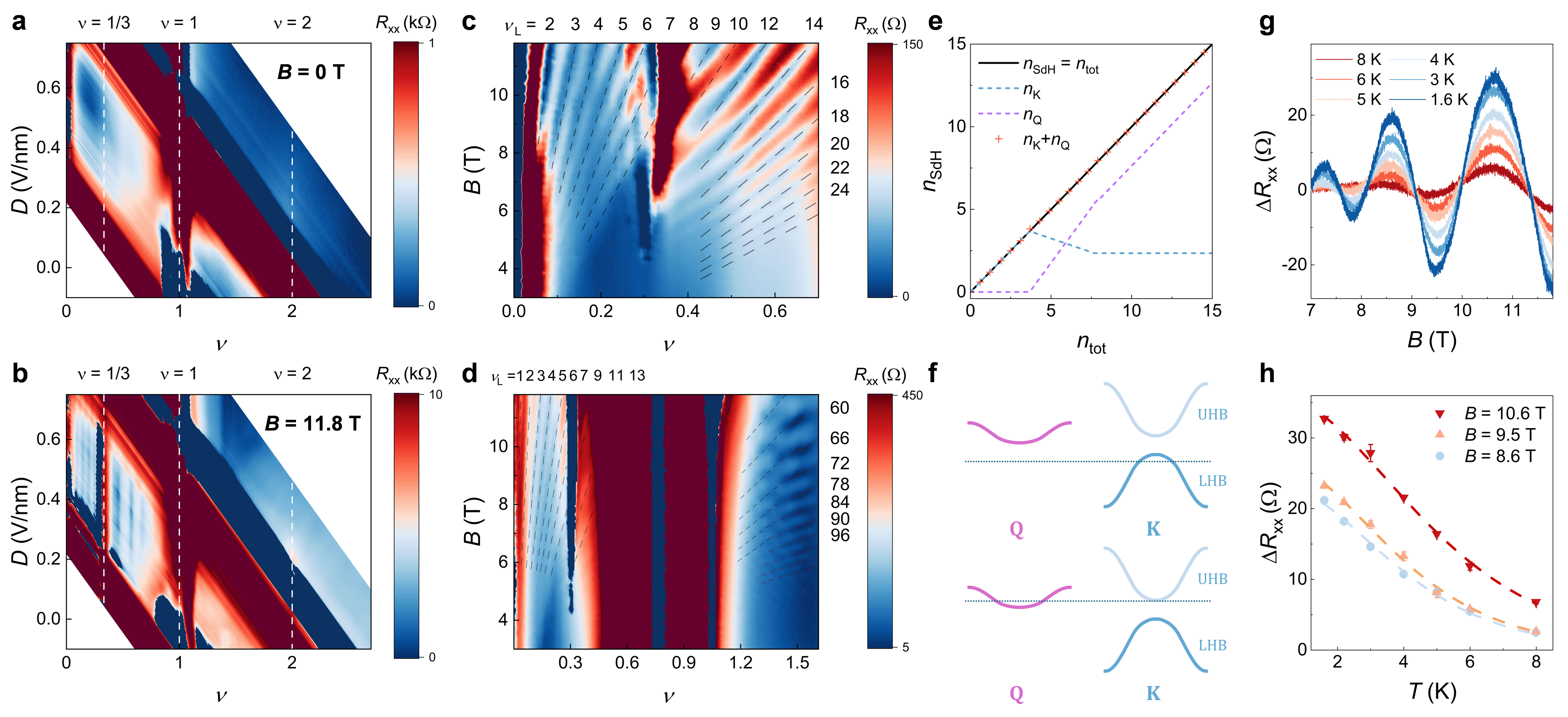}
\caption{Quantum oscillations in tMoS$_2$ (device D1). \textbf{a}, \textbf{b}, Color map of four-terminal resistance $R_{xx}$ as a function of $\nu$ and $D$ at $B = 0$ T (a) and $B = 11.8$ T (b) at 1.5 K. The filling factors of $\nu = 1/3$, $\nu = 1$, and $\nu = 2$ are marked. \textbf{c}, \textbf{d}, $R_{xx}$ as a function of $\nu$ and $B$ at fixed $D = 0.425$ V\,nm$^{-1}$ (c) and $D = 0.650$ V\,nm$^{-1}$ (d). \textbf{e}, Electron carrier density extracted from SdH oscillations ($n_{\text{SdH}}$) as a function of total carrier density induced by gates ($n_{\text{tot}}$). Black solid curve represents $n_{\text{SdH}} = n_{\text{tot}}$. The blue and purple dashed curves denote the electron densities inferred to be in the $K/K'$ bands $n_{K}$ and $Q/Q'$ bands $n_{Q}$, respectively (degeneracy $g = 1$ in $0 < n < 1.4 \times 10^{12}$ cm$^{-2}$, $g = 2$ in $n > 1.4 \times 10^{12}$ cm$^{-2}$ for $f_{K}$ and $g = 6$ for $f_{Q}$ are used based on $n = (ge/h) \times f$). Red crosses represent the summation of $n_{K}$ and $n_{Q}$. \textbf{f}, Schematic of the band alignment in the conduction band of tMoS$_2$. The dashed lines mark the Fermi levels, which are located at lower Hubbard band (LHB) and upper Hubbard band (UHB) in top panel and bottom panel, respectively. \textbf{g}, Temperature-dependent SdH oscillations at fixed $n = 1.1 \times 10^{12}$ cm$^{-2}$ and $D = 0.644$ V\,nm$^{-1}$. \textbf{h}, Fit of SdH oscillation amplitude $\Delta R_{xx}$ as a function of $T$ by L-K formula. The fit yields an effective mass $m^{*} = (0.35 \pm 0.01)m_{e}$.}
\label{fig:2}
\end{figure*}

The Landau fans measured at $D = 0.425$ and 0.650 V\,nm$^{-1}$ (Figs.~2c
and 2d, respectively) reveal onefold, twofold, and sixfold sequences of
Landau levels (LLs). The first two are compatible with isospin (spin-
and valley-) polarized and partially polarized states derived from the
$K/K'$ valleys [28--30]. The sixfold sequence indicates electrons
occupying the six unpolarized $Q$/$Q'$ valleys of MoS$_2$ [20,31,32].

To test this assignment, we extracted Shubnikov-de Haas (SdH)
oscillation density using $n_{\text{SdH}} = (ge/h)f$ as a function of
gate-induced density $n_{\text{tot}}$, where $f$ is the SdH oscillation
frequency, $g = g_{s}g_{\nu}$ accounts for spin and valley degeneracy,
$e$ is the elementary charge, and $h$ is the Planck's constant. As shown
in Fig.~2e, the density inferred using a onefold degeneracy below
$n_{\text{tot}} \approx 1.4 \times 10^{12}$ cm$^{-2}$ and twofold
degeneracy in the range of
$1.4 \times 10^{12}$ cm$^{-2} < n_{\text{tot}} < 3.8 \times 10^{12}$ cm$^{-2}$
agrees with the total density (see Fig.~S1 in Supplementary Material).
Above $n_{\text{tot}} \approx 3.8 \times 10^{12}$ cm$^{-2}$, a second
component emerges while the inferred $K/K'$ contribution starts to
decrease until $n_{\text{tot}} \approx 7.7 \times 10^{12}$ cm$^{-2}$.
Assigning the new component with a sixfold degeneracy yields a summation
of density $n_{K} + n_{Q}$ comparable to $n_{\text{tot}}$ (Fig.~2e).
These data support the band alignment schematic in Fig.~2f, in which the
$Q/Q'$ band becomes occupied before the upper Hubbard-like feature of the
$K/K'$ moir\'e band is reached (see additional supporting data in
Fig.~S2).

The multivalley nature of the moir\'e band is further evidenced by
comparing the effective mass $m^{*}$ extracted from the temperature
dependence of the SdH oscillations. The amplitude of the SdH
oscillations can be described by the Lifshitz-Kosevich (L-K) formula
[33,34]: $\Delta R_{xx} \propto \xi/\text{sinh}\xi$, where
$\xi = 2\pi^{2}k_{B}T/\hslash\omega_{c}$, $\omega_{c} = eB/m^{*}$,
$k_{B}$ is the Boltzmann constant, and $\hslash$ is the reduced Planck's
constant. By fitting the temperature-dependent SdH oscillations as shown
in Fig.~2g and Fig.~2h, an effective mass of approximately $0.35m_{e}$
can be obtained at $n = 1.1 \times 10^{12}$ cm$^{-2}$ and
$D = 0.644$ V\,nm$^{-1}$. In contrast, a larger effective mass of
$0.75m_{e}$ is extracted in the LLs region with a sixfold degeneracy
(see Fig.~S3 in Supplementary Material). The significantly heavier
$m^{*}$ at $Q/Q'$ valleys than that at $K/K'$ valleys is consistent with
previous theoretical calculations in few-layer MoS$_2$ without moir\'e
[35].

Notably, in intrinsic bilayer MoS$_2$, the measurement results indicate
only $K/K'$-valley-dominated transport behavior [29]. Our observation of
electron populations from the $Q/Q'$ valleys reveals that the band
folding induced by the moir\'e potential alters the band alignment
between $K/K'$ and $Q/Q'$ bands.

\subsection{Pomeranchuk Effect at $\nu = 1/3$}

We next turn to the correlated state at $\nu = 1/3$ in device D1. Figure
3a shows the evolution of $R_{xx}$ from 1.5 to 60 K. The fractional
resistance feature is barely discernible at base temperature but becomes
progressively more pronounced upon warming, and is strongest at
intermediate temperature. This anomalous evolution is also evident in
the fixed-$V_{bg}$ transfer curves in Fig.~3e. To isolate the fractional
feature from the broad gate-dependent background, we define a peak
magnitude $R^{*}$ by subtracting a smooth local background [36], as
illustrated in the inset of Fig.~3f. $R^{*}$ increases on warming to
approximately 13 K, decreases between 13 and 40 K, and is eventually
obscured by the high-temperature background. The nonmonotonic evolution
of the fractional resistance feature is distinct from the conventional
monotonic thermal melting of an insulating state.

\begin{figure*}[t]
\includegraphics[width=\textwidth]{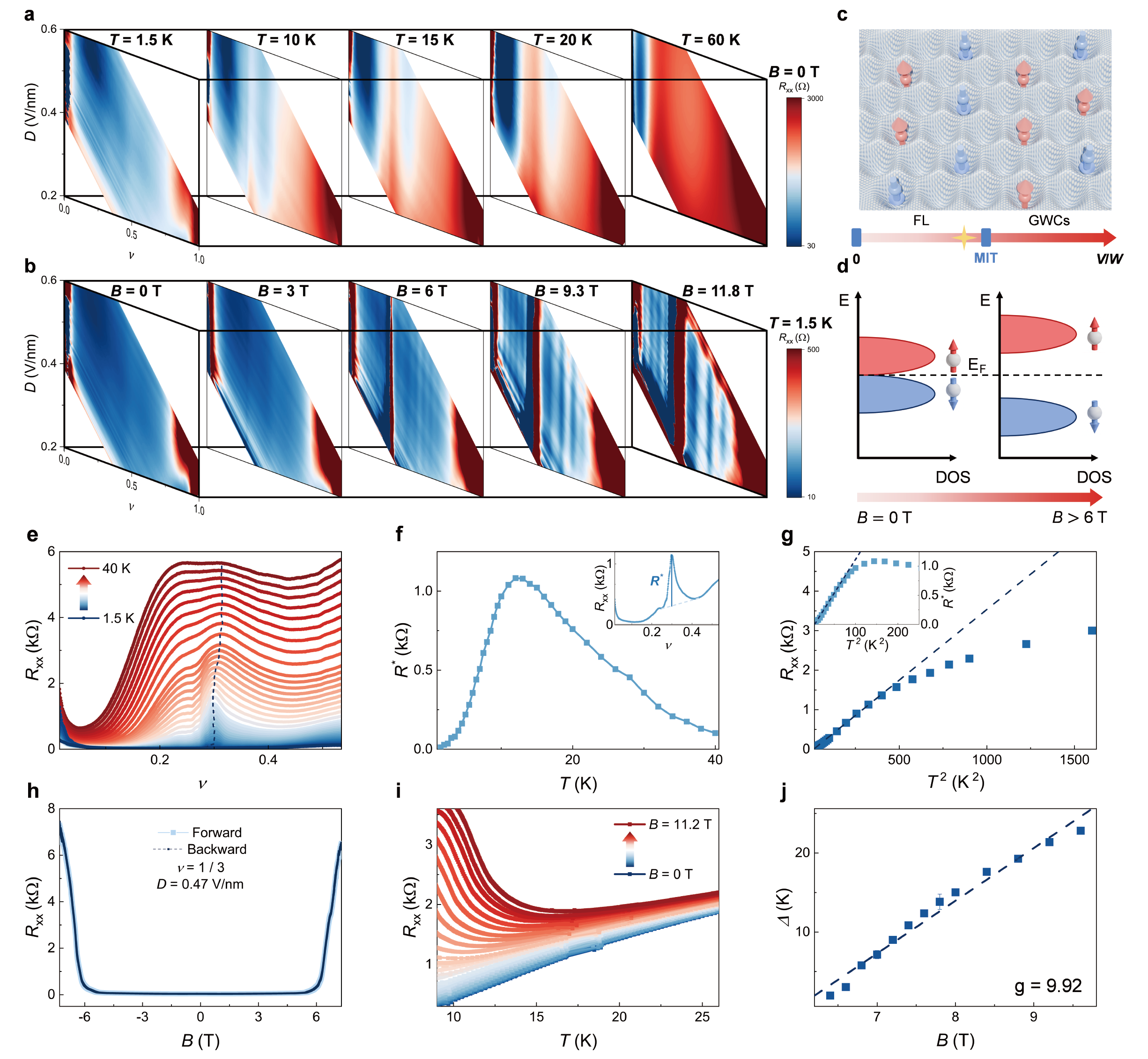}
\caption{Pomeranchuk effect at $\nu = 1/3$ in tMoS$_2$ (device D1). \textbf{a}, \textbf{b}, Evolution of $R_{xx}(\nu,D)$ maps near $\nu = 1/3$ as a function of $T$ (a) and $B$ (b). \textbf{c}, Up panel: Schematic of spatial charge distribution on the moir\'e superlattice at $\nu = 1/3$. Bottom panel: Schematic phase diagram of MIT transition at $\nu = 1/3$. The yellow star marks that our system lies on the left boundary of the MIT. \textbf{d}, Schematic of the magnetic field induced Zeeman splitting with gap opening. \textbf{e}, $R_{xx}$ as a function of $\nu$ at various temperature from 1.5 K to 40 K at fixed $V_{bg} = 5.9$ V. The dashed curve guides the shift of the correlated peaks. \textbf{f}, Temperature dependence of $R^{*}$, where $R^{*}$ is extracted by subtracting a smooth background (see the inset). \textbf{g}, $R_{xx}$ as a function of $T^{2}$ at low-temperature region ($T < 40$ K) at fixed $\nu = 1/3$ and $D = 0.384$ V\,nm$^{-1}$. The dashed line denotes the linear fit. The inset shows the corresponding fit using $R^{*}$. \textbf{h}, $R_{xx}$ as a function of $B$ at fixed $\nu = 1/3$ and $D = 0.470$ V\,nm$^{-1}$. \textbf{i}, $R_{xx}$ as a function of $T$ at fixed various $B$ at fixed $\nu = 1/3$ and $D = 0.470$ V\,nm$^{-1}$. \textbf{j}, Evolution of extracted gap $\Delta$ as a function of $B$. The dashed line shows the linear fit, yielding a g-factor $g^{*} \approx 9.92 \pm 0.47$.}
\label{fig:3}
\end{figure*}

At low temperature ($T < 9$ K), the transport at $\nu = 1/3$ follows a
standard Fermi-liquid behavior described by
$R_{xx}(T) = R_{0} + AT^{2}$ (Fig.~3g), where $R_{0}$ is the residual
resistance and $A \propto {(m^{*})}^{2}$ reflects the quasiparticle
effective mass. Such behavior confirms a metallic Fermi-liquid ground
state at base temperature. In this coherent liquid state, itinerant
electrons tend to form arranged spin pairs and therefore feature a low
entropy proportional to $T$ [37]. As the temperature rises
($9\text{~K} < T < 13\text{~K}$), $R_{xx}$ exhibits a pronounced
enhancement, signaling a temperature-driven electronic solidification,
namely, enhanced charge localization. At fractional moir\'e fillings,
inter-site Coulomb repulsion can overcome the residual kinetic energy and
stabilize commensurate charge-ordered states commonly described as GWC
[7,8,10]. In this high-temperature regime, thermal fluctuations suppress
the coherent electronic hopping, localizing electrons onto the lattice
sites. Due to the weak magnetic exchange interaction between these
spatially separated localized electrons, their isospin degrees
of freedom remain highly disordered owing to strong spin fluctuations or
geometric frustration, contributing a large, temperature-independent
residual entropy approaching $k_{B}\ln 2$ per electron [38]. This
behavior clearly demonstrates an electronic Pomeranchuk-like effect,
where charge localization is entropically driven by the highly degenerate
spin degrees of freedom.

Perpendicular magnetic field produces a similar enhancement of the
resistive states at $\nu = 1/3$, as shown in Fig.~3b, which plots the
$\nu-D$ map at fixed $B_{\bot}$. At fixed $\nu = 1/3$ and
$D = 0.470$ V\,nm$^{-1}$, the field sweeps in Fig.~3h reveal a
field-induced metal-insulator transition (MIT) with no detectable
hysteresis between the two sweep directions. This manifests as a
dramatic increase in $R_{xx}$ above a critical $B_{\bot} \approx 6$ T.
Consistently, the temperature dependences shown in Fig.~3i evolve from
metallic-like behavior at weak $B_{\bot}$ to insulating-like behavior at
larger $B_{\bot}$. Above $B_{\bot} > 6$ T, the insulating states follow
a thermal activation behavior described by
$R_{xx} \propto e^{\Delta/2k_{B}T}$, where $\Delta$ is the energy gap
(see Fig.~S5 in Supplementary Material). The insulating behavior
enhanced by magnetic fields can be considered as the result of Zeeman
effect [10]. Zeeman energy induces an energetic splitting of the
spin-unpolarized states by $E_{z} = g^{*}\mu_{B}B$, where $g^{*}$ is the
corresponding $g$-factor at $\nu = 1/3$. This $g$-factor can be
extracted from the evolution of $\Delta$ as a function of $B$ as shown
in Fig.~3j, from which we obtain a large $g^{*} \approx 9.92$. Under large
$B_{\bot}$, due to this large g-factor, Zeeman-field-induced spin
splitting facilitates the gap opening of undeveloped extended Hubbard
band at low temperature, as illustrated in Fig.~3d.

We find that the magnetoresistance under an in-plane magnetic field is
weak even in the large-field regime (see Supplementary Fig.~S4). Such
strong magnetic anisotropy is expected for spin-valley-locked electrons
in MoS$_2$ due to its strong Ising spin-orbit coupling.

Notably, this magnetic-field-induced MIT observed here exhibits similar
behavior to that observed at $\nu = 1$ in twisted MoTe$_2$/WSe$_2$
systems [16], where the Mott-Hubbard gap arises from on-site
(short-range) Coulomb repulsion interaction $U$ described by the
triangular-lattice Hubbard model. In contrast, our observations of MIT
at $\nu = 1/3$ in twisted MoS$_2$ systems reveal the competition between
extended Hubbard gaps and a spin-polarized band insulator. In our system,
the strong long-range Coulomb interaction $V$ dominates over $W$ with the
assistance of intermediate temperature or high $B_{\bot}$. Figure 3c
illustrates the real-space configurations of the formed GWC states, where
the filled electrons repel not only occupation of the same site, but also
occupation of the adjacent sites. As a result, the lattice translational
symmetry is spontaneously broken in our tMoS$_2$ system. The electrons
condense to a $\sqrt{3} \times \sqrt{3}$ super unit cell pattern.

With the observations of the correlated metal ground states at low
temperatures, the Pomeranchuk effect at intermediate temperatures, and
the magnetic-field induced MIT, we infer that the ground state in our
system lies in the vicinity of MIT point at intermediate coupling
strength as shown in the bottom panel of Fig.~3c.

\subsection{Pomeranchuk Effect at $\nu = 1/4$}

% Figure 4 float is defined here (immediately after the subsection header,
% encountered on page 5) so that it is placed at the top of page 6 rather
% than page 7.
\begin{figure*}[!t]
\includegraphics[width=\textwidth]{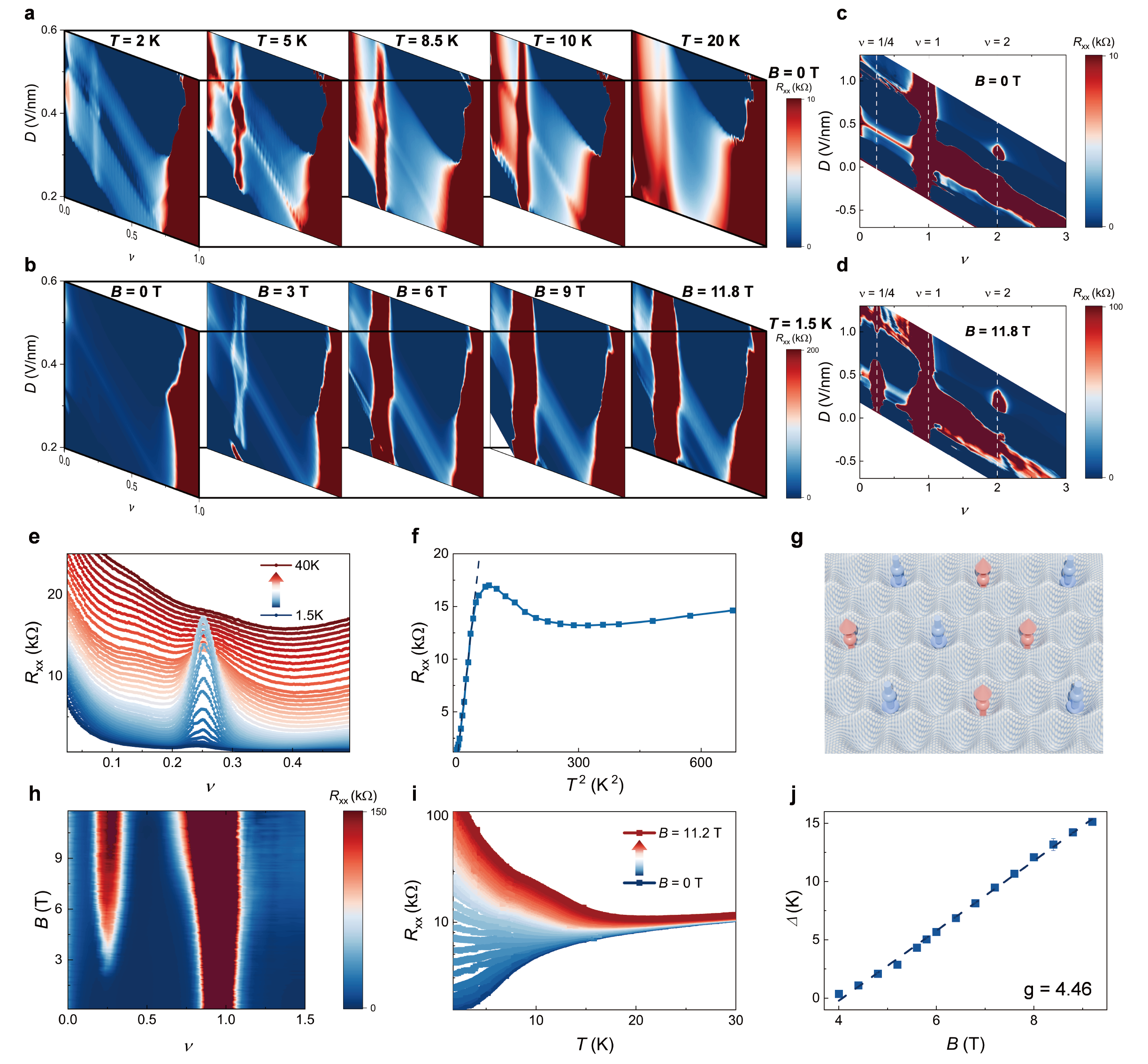}
\caption{Pomeranchuk effect at $\nu = 1/4$ in tMoS$_2$ (device D2). \textbf{a}, \textbf{b}, Evolution of $R_{xx}(\nu,D)$ maps near $\nu = 1/4$ as a function of $T$ (a) and $B$ (b). \textbf{c}, \textbf{d}, Color map of four-terminal resistance $R_{xx}$ as a function of $\nu$ and $D$ at $B = 0$ T (c) and $B = 11.8$ T (d) at 1.5 K. The dashed lines mark the position of $\nu = 1/4$, $\nu = 1$, and $\nu = 2$. \textbf{e}, $R_{xx}$ as a function of $\nu$ at various temperature from 1.5 K to 40 K at fixed $V_{bg} = 5.1$ V. \textbf{f}, $R_{xx}$ plotted as a function of $T^{2}$ at fixed $D = 0.350$ V\,nm$^{-1}$. The dashed line denotes the linear fit at low-temperature region. \textbf{g}, Schematic of spatial charge distribution on the moir\'e superlattice at $\nu = 1/4$. \textbf{h}, Color map of $R_{xx}$ as a function of $\nu$ and $B$ at fixed $V_{bg} = 5.0$ V. \textbf{i}, $R_{xx}$ as a function of $T$ at fixed various $B$ at fixed $\nu = 1/4$ and $D = 0.510$ V\,nm$^{-1}$. \textbf{j}, Evolution of extracted gap $\Delta$ as a function of $B$. The dashed line shows the linear fit, yielding a g-factor $g^{*} \approx 4.46 \pm 0.07$.}
\label{fig:4}
\end{figure*}

To test the reproducibility of Pomeranchuk effect at fractional
fillings, we have measured a second device D2 with a targeted twist
angle near $64^{\circ}$. The density at one electron per moir\'e unit
cell is $n_{M} \approx 5.2 \times 10^{12}$ cm$^{-2}$, corresponding to a
twist angle of ${3.9}^{\circ}$ for an AB-stacked tMoS$_2$. At 1.5 K, GWC
states emerge at $\nu = 1/4$ and are enhanced at $B_{\bot} = 11.8$ T as
the comparison of Figs.~4c and 4d. Its temperature evolution closely
parallels the behavior at $\nu = 1/3$ in device D1: the feature grows on
warming from base temperature, reaches a maximum near 10 K, and weakens
at higher temperature (Figs.~4a, 4e, and 4f). Meanwhile, a
$B_{\bot}$-induced MIT occurs above 3 T as shown in Fig.~4b, 4h, and 4i.
At low temperature, this state at $\nu = 1/4$ exhibits a typical
Fermi-liquid behavior following a $T^{2}$ dependence below approximately
6.5 K as shown in Fig.~4f. From the $R_{xx}(T)$ curves measured at
various $B$ shown in Fig.~4i, insulating gaps $\Delta$ above $4$ T are
fitted through Arrhenius plots. The evolution of $\Delta$ with $B$ shown
in Fig.~4j yields a $g$-factor of 4.46, lower than that at $\nu = 1/3$
in device D1.

The real-space electron configuration for the GWC state at $\nu = 1/4$
is illustrated in Fig.~4g, where spontaneous breaking of translational
symmetry in this device leads to the formation of stripe phases. The
change from $\nu = 1/3$ in device D1 with $\theta = {4.1}^{\circ}$ to
$\nu = 1/4$ in device D2 with $\theta = {3.9}^{\circ}$ indicates the
relative strength of $V/W$ varies with the moir\'e wavelength. As a
result, the inter-site Coulomb interaction may enable Wigner
crystallization at different fractional fillings of moir\'e unit cell.

\section{Conclusion}

In summary, we observe temperature- and magnetic-field-enhanced
localization at fractional moir\'e fillings in AB-stacked tMoS$_2$. These
results extend Pomeranchuk phenomenology to fractional moir\'e fillings
and identify tMoS$_2$ as a platform for entropy-driven electron
crystallization. At the fractional fillings of $\nu = 1/3$ and
$\nu = 1/4$, our data reveal common features: a low-temperature
Fermi-liquid-like $T^{2}$ regime, a finite-temperature strengthening of a
commensurate fractional resistance feature, and a $B_{\bot}$-induced
high-resistance regime. This phenomenology is consistent with a
competition between an itinerant state of relatively low entropy and a
localized GWC-like state that retains weakly coupled internal degrees of
freedom. In the latter state, reduced exchange can leave isospin moments comparatively disordered, providing an entropy
reservoir.

Our findings will motivate further experimental efforts to probe the
rich phenomenology involving multiple degrees of freedom in tMoS$_2$,
which would provide in-depth insights into the quantum phase transitions
between ordered and disordered correlated states in the
intermediate-correlation-strength regime. Direct thermodynamic probes
and local measurements will be important for determining the entropy and
charge-order pattern of this state [8,14].

\section{Methods}

\subsection{Device Fabrication}

Natural monolayer MoS$_2$ flakes were mechanically exfoliated and cut
into two pieces with a tungsten tip before dry transfer to form near-2H
twisted bilayers. The tMoS$_2$ was encapsulated by hBN sheets with
approximate top- and bottom-hBN thicknesses of 10-20 nm and 50-60 nm,
respectively. To obtain low temperature ohmic contacts, the top hBN was
pre-patterned and etched through windows of approximately
$1 \times 3$ $\mu$m$^{2}$ at the contact regions. Bi/Au electrodes and
the top gate were deposited after stack assembly. A large few-layer
graphite flake served as the bottom gate. The devices were finally
patterned into Hall bar geometries for simultaneous $R_{xx}$ and
$R_{xy}$ measurements.

Carrier density $n$ and displacement field $D$ were controlled
independently by the top- and bottom-gate voltages. We convert $V_{tg}$
and $V_{bg}$ into $n$ and $D$ by
$n = \frac{C_{tg}V_{tg} + C_{bg}V_{bg}}{e},\ D = \frac{C_{bg}V_{bg} - C_{tg}V_{tg}}{2\varepsilon_{0}}$,
where $C_{tg}$ and $C_{bg}$ are the corresponding gate capacitances per
unit area, respectively.

\subsection{Electrical Measurements}

All transport data were acquired in an Oxford TeslatronPT cryogen-free
refrigerator with a base temperature of 1.5 K and magnetic field up to
12 T. Measurements were performed using a standard low-frequency lock-in
technique at 17.777 Hz. The source-drain excitation current was 200 nA
for device D1 and 50 nA for device D2. Longitudinal and transverse
voltage drops were measured with voltage preamplifiers of 100
M$\Omega$ input impedance.

\begin{acknowledgments}
We thank Dr.~Chao Zhang from the Instrumentation and Service Center for
Physical Sciences (ISCPS) at Westlake University for technical support in
data acquisition. This work was funded by National Natural Science
Foundation of China (Grant No.~12574203, Grant No.~12550402, Grant
No.~12274354), the Zhejiang Provincial Natural Science Foundation of
China (Grant No.~LR24A040003), Hangzhou Natural Science Foundation for
Key Program (Grant No.~2025SZRJJ1093) and Westlake Education Foundation
at Westlake University. K.W. and T.T. acknowledge support from the JSPS
KAKENHI (Grant Numbers 21H05233 and 23H02052) and World Premier
International Research Center Initiative (WPI), MEXT, Japan. The authors
thank the ISCPS, the Instrumentation and Service Centers for Molecular
Science, and the Westlake Center for Micro/Nano Fabrication at Westlake
University for facility support.
\end{acknowledgments}

\end{document}